\documentclass[english,aps,floatfix]{revtex4}
\usepackage{graphicx}
\usepackage{dsfont}
\usepackage{enumerate}
\usepackage{slashed}
\usepackage{amsfonts}
\usepackage[dvips]{epsfig}
\usepackage{babel}
\usepackage[dvips]{color}
\usepackage{amssymb}

\usepackage[latin1]{inputenc}
\usepackage{amsmath}

\newtheorem{theorem}{Theorem}[section]

\newtheorem{e-proposition}[theorem]{Proposition}
\newtheorem{corollary}[theorem]{Corollary}
\newtheorem{e-definition}[theorem]{Definition\rm}

\begin{document}

\title{Study of a self-adjoint operator indicating the direction of time within standard quantum mechanics}

\author{Y. Strauss\textsuperscript{1}, J. Silman\textsuperscript{2}\footnote{Present address: Laboratoire d'Information Quantique, Universit\'{e} Libre de Bruxelles, 1050 Bruxelles, Belgium}, S. Machnes\textsuperscript{2}\footnote{Present address: Institut f\"{u}r Theoretische Physik, Universit\"{a}t
Ulm, 89069 Ulm, Germany}, L.P. Horwitz\textsuperscript{3, 4}}

\affiliation{\textsuperscript{1}Einstein Institute of Mathematics, Edmond J. Safra campus,
The Hebrew University of Jerusalem, Jerusalem 91904, Israel \\
\textsuperscript{2}School of Physics and Astronomy, Raymond and Beverly Sackler
Faculty of Exact Sciences, Tel-Aviv University, Tel-Aviv 69978, Israel \\
\textsuperscript{3}Physics Department of Physics, Bar-Ilan University, Ramat-Gan 52900, Israel \\
\textsuperscript{4}Department of Physics, The Ariel University Center of Samaria, Ariel 40700, Israel}

\begin{abstract}

\textbf{In [J. Math. Phys. 51 (2010) 022104] a self-adjoint operator was introduced that has the property that it indicates the direction of time within the framework of standard quantum mechanics, in the sense that
as a function of time its expectation value decreases monotonically for any initial state. In this paper we study some of this operator's properties. In particular, we derive its spectrum and generalized eigenstates, and treat the example of the free particle.}

\end{abstract}
\maketitle

\section{Introduction}

It is a fundamental question in standard quantum mechanics (SQM) of what type of restrictions the Schr\"{o}dinger evolution imposes on the behavior in time of basic objects. In particular, it is of interest to ask if SQM allows for self-adjoint operators having the so-called Lyapunov property, that is, monotonicity of the expectation value irrespective of the initial state of the system. Clearly, such an operator would indicate the direction of time.

A natural candidate for a self-adjoint Lyapunov operator is a (self-adjoint) time operator $T$ canonically conjugate to the Hamiltonian $H$, such that $T$ and $H$ form an imprimitivity system \cite{Mackey} (implying that each generates a translation on the spectrum of the other). However, a well known theorem of Pauli tells us that this is impossible \cite{Pauli}. Recently, Galapon attempted to bypass Pauli's arguments and found pairs of $T$ and $H$ satisfying the canonical commutation relations, but do not constitute an imprimitivity system \cite{Galapon}. It can be shown that the $T$ operator obtained in this way does not have the Lyapunov property. Other authors do not insist on the conjugacy of $T$ and $H$. In this context, Unruh and Wald's proof that a `monotonically perfect clock' does not exist \cite{Unruh} should be noted, as well as Misra, Prigogine, and Courbage's no-go theorem \cite{Misra}. Still another solution, advocated by some authors,  is to do away with the requirement of self-adjointness \cite{Holevo}. 

A Lypaunov self-adjoint operator acting within the framework of standard quantum mechanics was recently introduced in \cite{Yossi}.  In this paper we study some of its properties. In particular, we derive its spectrum and generalized eigenstates,  and treat the example of the free particle.

\section{Main Results}
Let $\mathcal H$ be a separable Hilbert space and let $H$ be a self-adjoint operator generating a unitary evolution group $\{U(t)\}_{t\in\mathbb R}$, with $U(t)=\exp(-iHt)$, on $\mathcal H$. We take $\mathcal H$ to represent the Hilbert space corresponding to some given quantum system and $H$ its Hamiltonian. For an initial state $\psi(0)=\psi\in\mathcal H$ $\psi(t)=U(t)\psi$ denotes the state of the system at time $t$ and $\Psi_\psi:=\{\psi(t)\}_{t\in\mathbb R^+}$ its trajectory. Let $\mathcal B(\mathcal H)$ be the space of bounded linear operators on $\mathcal H$. A (forward) Lyapunov operator is defined as follows \cite{Yossi}:
\begin{e-definition} 
Let $M\in\mathcal B(\mathcal H)$ be a self-adjoint operator on $\mathcal H$. Let $\Psi_\psi$ be the trajectory corresponding to an initial state $\psi$. Let $M(\Psi_\psi)=\{\Vert\varphi\Vert^{-2}(\varphi,\,M\varphi)\mid \varphi\in\Psi_\psi\}$ be the set of expectation values of $M$ in states in $\Psi_\psi$. Then $M$ is a forward Lyapunov operator if the mapping $\tau_{M,\,\psi}:\mathbb R^+\mapsto 
M(\Psi_\psi)$ defined by
\begin{equation}
 \tau_{M,\,\psi}(t)=\Vert\psi(t)\Vert^{-2}(\psi(t),\,M\psi(t))=\Vert\psi\Vert^{-2}(\psi(t),\,M\psi(t))
\end{equation}
is one to one and monotonically decreasing for all non-recurring trajectories.\hfill$\square$
\end{e-definition}

\par Let $\mathcal K$ be a separable Hilbert space and $L^2(\mathbb R^+;\,\mathcal K)$ the Hilbert space of Lebesgue 
square-integrable $\mathcal K$ valued functions defined on $\mathbb R^+$, let $H$ be the operator of multiplication by the 
independent variable on $L^2(\mathbb R^+;\,\mathcal K)$, and let $\{U(t)\}_{t\in\mathbb R}$ be the continuous one-parameter 
unitary evolution group generated by $H$, i.e.,
\begin{equation}
\label{eq:L_2_schrodinger_evol}
 [U(t)f](E)=[e^{-iHt}f](E)=e^{-iEt}f(E),\quad f\in L^2(\mathbb R^+;\,\mathcal K),\ \ E\in\mathbb R^+.
\end{equation}
Denote by $\mathcal H^2(\mathbb C^\pm;\,\mathcal K)$ the Hardy spaces of 
$\mathcal K$ valued functions analytic in $\mathbb C^\pm$.
 The Hilbert spaces $\mathcal H^2_\pm(\mathbb R;\,\mathcal K)$ consisting of non-tangential boundary 
values on $\mathbb R$ in $\mathcal H^2(\mathbb C^\pm;\,\mathcal K)$ are, respectively, isomorphic to $\mathcal H^2(\mathbb 
C^\pm;\,\mathcal K)$. The spaces 
$\mathcal H^2_\pm(\mathbb R;\,\mathcal K)$ are orthogonal subspaces of $L^2(\mathbb R;\,\mathcal K)$ with
$L^2(\mathbb R;\,\mathcal K)=\mathcal H^2_+(\mathbb R;\,\mathcal K)\oplus\mathcal H^2_-(\mathbb R;\,\mathcal K)$.
We denote the orthogonal projections in $L^2(\mathbb R;\,\mathcal K)$ on $\mathcal H^2_+(\mathbb R;\,\mathcal K)$ and $\mathcal 
H^2_-(\mathbb R;\,\mathcal K)$, respectively, by $P_+$ and $P_-$. Let $L^2(\mathbb R^\pm;\,\mathcal K)$ be the subspaces of $L^2(\mathbb R;\,\mathcal K)$ consisting of square-integrable functions supported on 
$\mathbb R^\pm$. There exists another orthogonal decomposition $L^2(\mathbb R;\,\mathcal K)=L^2(\mathbb R^-;\,\mathcal K)\oplus L^2(\mathbb R^+;\,\mathcal K)$ with projections on $P_{\mathbb R^+}$ and $P_{\mathbb R^-}$. The following theorem is proved in \cite{Yossi}:
\begin{theorem}
\label{thm:L_2_lyapunov_op}
Let $M: L^2(\mathbb R^+;\,\mathcal K)\mapsto L^2(\mathbb R^+;\,\mathcal K)$ be the operator defined by
\begin{equation}
\label{eq:L_2_M}
 M:=(P_{\mathbb R^+}P_+P_{\mathbb R^+})\vert_{L^2(\mathbb R^+;\,\mathcal K)}\,.
\end{equation}
Then $M$ is a positive, contractive and injective operator on $L^2(\mathbb R^+;\,\mathcal K)$, such that $\text{Ran}\, M$ is 
dense in $L^2(\mathbb R^+;\,\mathcal K)$ and $M$ is a forward Lyapunov operator for the evolution on 
$L^2(\mathbb R^+;\,\mathcal K)$ defined in Eq. (\ref{eq:L_2_schrodinger_evol}), i.e., for every 
$\psi\in L^2(\mathbb R^+;\,\mathcal K)$ we have
\begin{equation}
 (\psi(t_2),\,M\psi(t_2))\leq (\psi(t_1),\,M\psi(t_1))\,,\quad t_2\geq t_1\geq 0\,,\quad \psi(t)=U(t)\psi\,,
\end{equation}
and moreover $ \lim_{t\to\infty}(\psi(t),\,M\psi(t))=0$. \hfill$\square$
\end{theorem}
\par The following corollary to Theorem \ref{thm:L_2_lyapunov_op} casts Eq. (\ref{eq:L_2_M}) in a more useful form:
\begin{corollary}
\label{cor:L_2_lyapunov_op}
Let $M$ be the Lyapunov operator given in Eq. (\ref{eq:L_2_M}). Then, for any function $f\in 
L^2(\mathbb R^+;\,\mathcal K)$ we have
\begin{equation}
\label{eq:M_explicit}
 (M f)(E)= -\frac{1}{2\pi i}\int_0^\infty dE'\frac{1}{E-E'+i0^+}f(E')\,,\quad E\in\mathbb R^+\,,\ \ 
 f\in L^2(\mathbb R^+;\,\mathcal K)\,.
\end{equation}
\hfill$\square$
\end{corollary}
{\bf Proof:} For every function $g\in\mathcal H^2(\mathbb C^\pm;\,\mathcal K)$ one has \cite{Paley-Wiener} 
\begin{equation}
 \mp\frac{1}{2\pi i}\int_{-\infty}^{\infty}dt\frac{1}{z-t}g_\pm(t)=
 \begin{cases} g(z)\,,& \pm \mathrm{Im}\,z>0\\ 0\,, & \pm \mathrm{Im}\, z<0\\ \end{cases}\,,
\end{equation}
where $g_\pm\in\mathcal H^2_\pm(\mathbb R;\,\mathcal K)$ is the boundary value function of $g$ on $\mathbb R$. Hence, the orthogonal projection $P_+$ of $L^2(\mathbb R^+;\,\mathcal K)$ on $\mathcal H^2_+(\mathbb R;\,\mathcal K)$ is explicitly given  by
\begin{equation}
\label{eq:P_plus_explicit}
 (P_+g)(\sigma)=-\frac{1}{2\pi i}\int_{-\infty}^\infty d\sigma'\frac{1}{\sigma-\sigma'+i0^+}g(\sigma')\,,
 \quad g\in L^2(\mathbb R;\,\mathcal K)\,,
\end{equation}
where Eq. (\ref{eq:P_plus_explicit}) hold a.e. for $\sigma\in\mathbb R$. From Eqns. (\ref{eq:L_2_M}), 
(\ref{eq:P_plus_explicit}) we then immediately obtain Eq. (\ref{eq:M_explicit}).\hfill$\blacksquare$
\begin{theorem}
\label{thm:M_spectrum}
The spectrum of $M$ satisfies $\sigma(M)=\sigma_{ac}(M)=[0,\,1]$. In $L^2(\mathbb R^+;\,\mathcal K)$ choose an orthogonal basis 
$\{\mathbf e_\lambda(E)\}_{E\in\mathbb R^+;\,\lambda\in\Lambda}$ with $\Lambda$ an appropriate index set such that for every 
$E\in\mathbb R^+$ the set $\{\mathbf e_\lambda(E)\}_{\lambda\in\Lambda}$ is a basis of $\mathcal K$ and we have 
$(\mathbf e_\lambda(E),\mathbf e_{\lambda'}(E'))_{L^2(\mathbb R^+;\,\mathcal K)}=\delta(E-E')\delta_{\lambda\lambda'}$, 
where $\delta_{\lambda\lambda'}$ stands for the Kronecker delta for discrete indices and the Dirac delta for continuous indices. Then 
for $m\in(0,\,1)$ the function defined by
\begin{equation}
\label{eq:M_eigenfunctions}
 \mathbf g_{m,\,\lambda}=\int_{0}^{\infty}dE\frac{E^{-\frac{i}{2\pi}\ln\left(\frac{1-m}{m}\right)
 -\frac{1}{2}}}{2\pi\sqrt{m\left(1-m\right)}}\mathbf e_\lambda(E)\,
\end{equation}
is a generalized eigenfunction of $M$ satisfying $M\,\mathbf g_{m,\,\lambda}=m\,\mathbf g_{m,\,\lambda}$. These eigenfunctions 
are normalized in such a way that 
$(\mathbf g_{m,\,\lambda},\mathbf g_{m',\,\lambda'})_{L^2(\mathbb R^+;\,\mathcal K)}=\delta(m-m')\delta_{\lambda\lambda'}$ and we 
have the eigenfunction expansion
\begin{equation}
 M=\sum_{\lambda\in\Lambda}\int_0^1 dm\,m\,\mathbf g_{m,\,\lambda}\,\mathbf g_{m,\,\lambda}^*\,.
\end{equation}
\hfill$\square$
\end{theorem}
{\bf Proof:} Consider the eigenvalue equation $M\, \mathbf g_{m,\,\lambda}=m\,\mathbf g_{m,\,\lambda}$. In the basis \linebreak
$\{\mathbf e_\lambda(E)\}_{E\in\mathbb R^+;\,\lambda\in\Lambda}$ the kernel of $M$ is given by 
$(\mathbf e_\lambda(E),M\,\mathbf e_{\lambda'}(E'))= -(2\pi i)^{-1}(E-E'+i0^+)^{-1}\delta_{\lambda\lambda'}$. Hence, the 
eigenvalue equation for $M$ assumes the form
\begin{equation}
\label{eq:eigenvalue_equation}
 -\frac{1}{2\pi i}\int_{0}^{\infty}dE'\frac{1}{E-E'+i0^+}\,g_{m,\,\lambda}\left(E'\right)
  =m\,g_{m,\,\lambda}\left(E\right)
 \,,\qquad E\in\mathbb R^+\,.
\end{equation}
where $g_{m,\,\lambda}(E)=(\mathbf e_\lambda(E),\mathbf g_{m,\,\lambda})$. 
Assume that $m\in(0,\,1)$.  To solve Eq. (\ref{eq:eigenvalue_equation}) we assume that the function $g_{m,\,\lambda}(E)$ 
is the boundary value on $\mathbb R^+$ from above of an analytic function $\tilde g_m(z)$ defined in 
$\mathbb C\backslash \mathbb R^+$. So that $g_{m,\,\lambda}(E)=\tilde g_{m,\,\lambda}(E+i0^+)$. 
This allows us to analytically continue eq. (\ref{eq:eigenvalue_equation}) into the cut complex plane 
\begin{equation}
\label{eq:complex_eigenvalue_eqn}
 -\frac{1}{2\pi i}\int_{0}^{\infty}dE'\frac{1}{z-E'}\,g_{m,\,\lambda}\left(E'\right)=m\,\tilde g_{m,\,\lambda}\left(z\right)
 \,,\qquad\mathrm{Im}z\neq 0\,.
\end{equation}
Taking in Eq. (\ref{eq:complex_eigenvalue_eqn}) the difference between the limits from above and below 
$\mathbb R^+$ we get 
\begin{equation}
\label{eq:limit_difference_II}
 g_{m,\,\lambda}\left(E\right)=\tilde g_{m,\,\lambda}(E+i0^+)=m\left(\tilde g_{m,\,\lambda}(E+i0^+)-\tilde g_{m,\,\lambda} 
 (E-i0^+)\right)\,.
\end{equation}
The function $\tilde g_{m,\,\lambda}\left(z\right)$ can now be continued to a second Riemann sheet by making 
use of the branch cut along $\left[0,\,\infty\right)$ in eq. (\ref{eq:complex_eigenvalue_eqn}). Denoting this two sheeted function again by $\tilde g_{m,\,\lambda}(z)$, 
Eq. (\ref{eq:complex_eigenvalue_eqn}), reduces to
\begin{equation}
\label{eq:riemann_surface_eigenvalue_eqn}
 \tilde g_{m,\,\lambda}(e^{2\pi i}z)=-\frac{1-m}{m}\, \tilde g_{m,\,\lambda}\left(z\right)\,.
\end{equation}
Eq. (\ref{eq:riemann_surface_eigenvalue_eqn}) admits solutions of the form $\tilde g_{m,\,\lambda}\left(z\right)=N_{m}z^{\beta}$
with $\beta=k+\frac{1}{2}-\frac{i}{2\pi}\ln\left(\frac{1-m}{m}\right)$, $k\in\mathbb{Z}$, and $N_{m}$ a normalization factor 
dependent on $m$. Setting $k=-1$ and $N_{m}=\left(4\pi^2 m\left(1-m\right)\right)^{-1/2}$ the solutions of the eigenvalue 
equation, Eq. (\ref{eq:eigenvalue_equation}), satisfy 
$\int_{0}^{\infty}dE\,\overline{g_{m',\,\lambda'}\left(E\right)}g_{m,\,\lambda}\left(E\right)
=\delta\left(m-m'\right)\delta_{\lambda\lambda'}$. With this choice of $k$ and 
$N_m$ the generalized eigenfunctions of $M$ are given by Eq. (\ref{eq:M_eigenfunctions}). 
\par We proceed to show by construction that the set of generalized eigenfunctions, Eq. (\ref{eq:M_eigenfunctions}), forms a 
complete set for the operator $M$. 
Consider the following integral

\begin{equation}
 \sum_\lambda\int_0^1 dm\,m\,\mathbf g_{m,\,\lambda}\,\mathbf g_{m,\,\lambda}^*=\sum_\lambda\int_{0}^{\infty}dE\int_{0}^{\infty}dE'\,
 \frac{\mathbf e_\lambda(E)\mathbf e^*_\lambda(E')}{4\pi^2\sqrt{E\,E'}}\int_0^1 \frac{dm}{1-m}\, \left(\frac{m}{1-m}\right)^{\frac{i}{2\pi}\ln (E/E')}.\label{eq:M_spectral_representation}
\end{equation}  
where $dm$ is a Lebesgue measure on $[0,\,1]$. The right hand side of eq. (\ref{eq:M_spectral_representation}) is readily obtained from that of eq. (\ref{eq:M_eigenfunctions}). Recall the definition of the Euler beta function $B(x,y)=\int_0^1 
dm\,m^{x-1}(1-m)^{y-1}=\Gamma(x)\Gamma(y)\Gamma^{-1}(x+y)$. $B(x,y)$ is well defined for $\mathrm{Re}\, x>0$ and 
$\mathrm{Re}\, y>0$, and can be analytically continued to other parts of the complex $x$ and $y$ planes. However, it 
is not well defined for $x=0$ and $y=0$, i.e. whenever $E=E'$ in Eq. (\ref{eq:M_spectral_representation}). To 
avoid this problem we shift $E$ (or $E'$) away from the real axis. Hence, in the integration over $m$ on the right hand side of 
Eq. (\ref{eq:M_spectral_representation}) we take the limit
\begin{eqnarray}
\label{eq:M_spectral_representation_II}
 & \frac{1}{4\pi^2}(E\,E')^{-1/2}\lim_{\theta\to 0^+}
 \int_0^1 dm\, (1-m)^{-\frac{i}{2\pi}\ln({e^{i\theta}E/E')-1}}m^{\frac{i}{2\pi}\ln(e^{i\theta}E/E')} &\nonumber\\
 &=\frac{1}{4\pi^2}(E\,E')^{-1/2}\lim_{\theta\to 0^+}
 B\Bigl(1-\frac{\theta}{2\pi}+\frac{i}{2\pi}\ln\bigl(\frac{E}{E'}\bigr)\ ,\,\frac{\theta}{2\pi}-\frac{i}{2\pi}\ln\bigl(\frac{E}{E'}\bigr)\Bigr)&\nonumber\\
 &=\frac{1}{4\pi^2}(E\,E')^{-1/2}\lim_{\theta\to 0^+}
 \frac{\pi}{\sin\left(-\frac{i}{2\pi}\ln\left(\frac{e^{i\theta} E }{E'}\right)\right)} = -\frac{1}{2\pi i} \frac{1}{E-E'+i0^+} &\,.
\end{eqnarray}
where in the last line in Eq. (\ref{eq:M_spectral_representation_II}) we used the identity 
$\Gamma(z)\Gamma(1-z)=\pi\sin^{-1}(\pi z)$.
Thus, the intgral on the spectrum of $M$ on the left hand side of Eq. (\ref{eq:M_spectral_representation_II}) reconstructs the 
kernel of $M$ in Eq. (\ref{eq:M_explicit}), i.e., we have reconstructed the operator $M$ using the set of generalized 
eigenfunctions $\{\mathbf g_{m,\,\lambda}\}_{m\in(0,\,1),\,\lambda\in\Lambda}$. Taken together with 
Eq. (\ref{eq:M_spectral_representation}), Eq. (\ref{eq:M_spectral_representation_II}) also shows 
that $M$ has no point spectrum at $m=0$ and $m=1$ and that $\sigma(M)=\sigma_{ac}(M)=[0,\,1]$. \hfill$\blacksquare$

\par Consider a quantum mechanical problem for which the Hamiltonian $H$, defined on an appropriate separable Hilbert space 
$\mathcal H$, satisfies the conditions: (i) Its ac spectrum is $\sigma_{ac}(H)=\mathbb R^+$. (ii) 
The multiplicity of $\sigma_{ac}(H)$ is uniform on $\mathbb R^+$. Let $\mathcal H_{ac}$ be the subspace of $\mathcal H$ corresponding to the 
ac spectrum of $H$. Then 
$\mathcal H_{ac}$ has a representation in terms of a function space $L^2(\mathbb R^+;\,\mathcal K)$, where $\mathcal K$ is a 
Hilbert space whose dimension correpsonds to the multiplicity of $\sigma_{ac}(H)$ and the evolution generated by $H$ is 
represented by  $\{U(t)\}_{t\in\mathbb R}$ defined in 
Eq. (\ref{eq:L_2_schrodinger_evol}). For such models we can apply the results discussed above. Using the standard Dirac 
notation, Eq. (\ref{eq:L_2_schrodinger_evol}) for $M$ can be rewritten as

\begin{equation}
 M = -\frac{1}{2\pi i}\sum_\lambda\int_0^\infty dE\int_0^\infty dE' 
 \vert E,\,\lambda\rangle\frac{1}{E-E'+i0^+}\langle E',\,\lambda\vert\,,\label{Dirac representation}
\end{equation}   
where the summation over $\lambda$ stands for summation 
over discrete degeneracy indices and appropriate integration over continuous degeneracy indices.
%

\begin{figure}{t}
\center{ \includegraphics[scale=0.6]{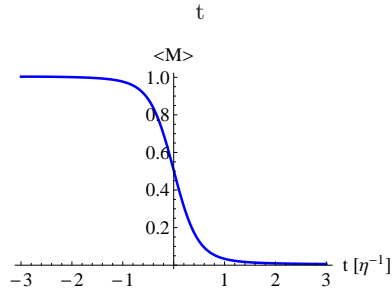}

\caption{Monotonic decrease of the expectation value of $M$ for a free Gaussian wave-packet with $p_{0}=0.64\eta$ and $\xi_{0}=0.3\eta$.}

}
\end{figure}

\par As an example we take
a one-dimensional Gaussian wave-packet representing the propagation
free particle of mass $\eta$ along $\mathbb R^+$  \begin{equation}
\psi\left(x,\, t\right)=\left(\frac{\eta^{2}\xi_{0}^{2}}{\pi\left(\eta+i\xi_{0}^{2}t\right)^{2}}\right)^{1/4}\exp\left(-\frac{\eta\xi_{0}^{2}x^{2}+ip_{0}\left(p_{0}t-2\eta x\right)}{2\left(\eta+i\xi_{0}^{2}t\right)}\right)\,.\label{1D Gaussian wave-packet}\end{equation}
 $p_{0}$ and $\xi_{0}$ are the location and width of the wave-packet
in momentum space at $t=0$. After obtaining the generalized eigenfunctions
of the free Hamiltonian, constructing the Lyapunov operator $M$ according
to Eqs. (\ref{Dirac representation}) and (\ref{eq:M_spectral_representation}), and applying it to the wave-packet
in Eq. (\ref{1D Gaussian wave-packet}) we obtain Fig. 1, showing
the time evolution of the expectation value of $M$, and Fig. 2, showing the time evolution of the associated probability
density function in the position basis and that of the eigenstates of $M$.
We see that if the sequence of time frames is shown in reverse order one is unable to tell whether
time is running backwards or whether one is observing a Gaussian wave-packet
propagating along the $\mathbb R^-$ (with time running forward). However,
if to all frames we attach the expectation value of $M$, then it possible to distinguish between these two
scenarios. The example plainly illustrates the time-ordering
of states introduced into the Hilbert space by the existence of a
Lyapunov operator.


\begin{figure}[h]
\center{ \includegraphics[scale=0.35]{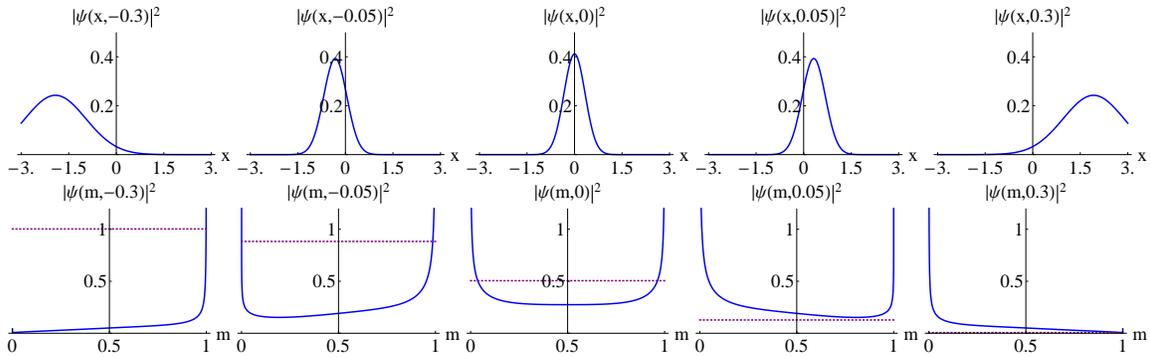}

\caption{Time frames of $\vert\psi (x,\,t)\vert^2$ and $\vert\langle m\mid \psi(t)\rangle\vert^2$ for the Gaussian wave-packet of Fig. 1.}

}
\end{figure}

In a forthcoming publication it will be shown that the existence
of the operator $M$ leads to an irreversible representation of the quantum
dynamics, and to a novel representation of the scattering process, in which the contribution of the resonance is singled out. \\

\emph{Addendum} -- This paper presents a mathematically rigorous reworking of some of the results appearing in \cite{arrow}, in particular, the derivation of the operator's spectrum and generalized eigenfunctions. The operator has since been generalized to a larger class of self-adjoint Lyapunov operators \cite{hegerfeldt}.

\section*{Acknowledgements}

Y.S. acknowledges the support of 
the Israeli Science Foundation Grant no. 1169/06, and the Center for Advanced Studies in Mathematics at Ben-Gurion University, where part of this research was conducted. J.S. and S.M. acknowledge the support of the Israeli Science Foundation Grant no. 784/06.

\end{document}